\documentclass{PoS}
\title{Nearby radio loud AGN and the Unified Model}

\ShortTitle{Nearby radio loud AGN and the Unified Model}

\author{\speaker{Elisabetta Liuzzo}
         \thanks{member of IT ARC node - Bologna (Italy)}\\
        INAF-Istituto di Radioastronomia \\
        E-mail: \email{liuzzo@ira.inaf.it}}

\author{Gabriele Giovannini\\
       Dipartimento di Fisica ed Astronomia, Bologna\\
       INAF-Istituto di Radioastronomia
       }

\author{Marcello Giroletti\\
       INAF-Istituto di Radioastronomia\\
       }

\abstract{The statistical study of the parsec scale properties of radio sources is crucial to get information on the nature of the central engine and to provide the
foundations of the current unified theories, suggesting that the appearance of active galactic nuclei depends strongly on orientation.
We started a project to observe at sub-arcsec resolution a complete sample of 94 nearby (z<0.1) radio galaxies, the Bologna Complete Sample, which is not affected by any selection
effect on the jet velocity and orientation with respect to the line of sight. Up to now, we published our parsec scale analysis of 77/94 sources. Here, we describe the last VLBA observations at 5 GHz and EVN data at 18 cm obtained for the 17 remaining faintest radio core ($<$ 5 mJy at 5 GHz in VLA images) BCS sources and we report our preliminary results on the whole complete sample.}

\FullConference{12th European VLBI Network Symposium and Users Meeting\\
                7-10 October 2014\\
                Cagliari, Italy}

\begin{document}

\section{Introduction}
The study of the parsec-scale properties of radio sources is crucial to obtain information on the
nature of the central engine and provide the foundations for the current unified theories, which
suggest that the appearance of active galactic nuclei (AGN) depends strongly on orientation.
The standard unified scheme is nowadays generally accepted as a consequence of relativistic
beaming effects and obscuration related to the orientation with respect to the line of sight. However, many
recent findings are posing challenges for the standard unified scheme. The discovery of
quasars with a FR I morphology, BL Lac objects with broad emission lines and high radio power (in the
FR II range) suggests that we still have to improve our knowledge in this field. Recently,
Kharb et al. 2009 found that among MOJAVE Blazars many BL Lacs exhibit radio
power and kpc scale morphology typical of FR II sources, while a substantial number of quasars
show a radio power intermediate between FR I and FR II. Moreover, the Fermi satellite has detected
gamma-ray emission in FR I radio galaxies and in steep spectrum radio sources. The origin and
nature of this high frequency emission is still under discussion. At the same time, the number of
radio sources with a very low core radio power in complete samples is high, suggesting the presence
of both high and low activity regimes in nuclear sources, and the possibility of restarted activity.
However, the duty cycle, and the triggering mechanisms, are not yet understood.
To properly discuss these points and to improve our knowledge of AGN, observations of statistical properties a large sample of radio sources unbiased 
with respect to orientation effects are
needed. Large surveys such as the Caltech-Jodrell Bank survey (e.g., Taylor et al. 1994) observed
sources selected at high frequency and therefore biased towards objects at small angles to the line
of sight. Large samples of well studied sources such as the MOJAVE sample collect only sources
with a bright radio core or gamma-ray emission, and are also likely biased to sources at small angles
with respect to the line of sight.
At present the largest complete sample with a large amount of available data on the kpc and
pc scale is the Bologna Complete Sample (BCS) which we are studying in detail over several years 
(Giovannini et al. 2001, 2005, and Liuzzo et al. 2009a, 2009b). It consists of 94 sources selected at
low frequency including all the B2 and 3CR radio sources present with z < 0.1, regardless of the core
flux density. Therefore, it is a sample not affected by any selection effect on the jet velocity and orientation
with respect to the line of sight.
For all the galaxies in our sample good arcsecond scale radio maps, as well X-ray data and HST optical images are available, allowing the necessary the comparison between the parsec and kiloparsec-scale radio structure and multiband emission properties. Up to now, we published results obtained at parsec scale for 77 of the 94 sources using VLBI observations (Giovannini et al. 2001, 2005, and Liuzzo et al. 2009a, 2009b).

\section{New VLBI data}
To complete the parsec scale analysis of the BCS, we asked and obtained new multifrequency VLBI data of the 17 remaining sources with very faint radio core ($<$ 5 mJy in 5 GHz VLA data) not yet observed in radio band at this high resolution. Thanks to these data, we would like to clarify some points as: a) how many two-sided sources are present in this complete sample? Are there non-relativistic or only mildly relativistic parsec-scale jets? Is the jet
velocity related to the core radio power? b) how to explain the faint radio core with the extended kpc scale structure and LogP(tot,408MHz)$>$ 25.0 W/Hz
observed in 7/17 sources of these very faint objects? c) Could we give constraints on the duty cycle of radio loud AGN on the base of statistical considerations of the BCS? d) what can we understand of these faint peculiar sources compared to the high-power ones?\\
In particular, we asked and obtained phase referencing observations with: \\
1)  {\bf VLBA at 5 GHz} ($\sim$ 2h on source at 512 Mbps) for homogeneity with observations of the others BCS sources which also show
that many objects have a nuclear emission self-absorbed at 5-8GHz.\\
2) {\bf EVN at 18 cm} ($\sim$ 2h on source at 1 Gbps) to compare with 5 GHz VLBA data and derive spectral index information for these sources showing a non dominant core emission. Moreover, the good (u,v) coverage of these data at the short baselines could allow in principle to map the possible intermediate scale structures poorly known in active radiogalaxies.\\
The achieved angular resolution in the final images is $\sim$ 15 $\times$ 5 mas$^{2}$ for EVN data while $\sim$ 3 $\times$ 2 mas$^{2}$ for VLBA data. The noise level is $\sim$ 0.03-0.12 mJy/beam for EVN data and $\sim$ 0.05-0.13 mJy/beam for VLBA data . The detection rate is 14/17 detected with S(tot) $<$ 4 mJy both at 1.7 GHz (EVN) and 5 GHz (VLBA). Most of these very faint sources are point-like, i.e. no jet clearly visible from modelfit results (e.g. see Fig.1). The core dominance (CD) distribution for the 17 very faint BCS sources shows that 12/17 radiogalaxies have CD between 0.25 and 1, while 5/17 objects have CD values $<$ 0.25 suggesting that nuclear variability and greater radio activity are present in the past.
\section{Preliminary results on the complete sample}
The new VLBI data of these 17 very faint objects allow to complete the pc-scale analysis of the BCS sample. In the following, we report our preliminary findings considering all the 94 sources of the sample.\\
a) The VLBI detection rate is high (93\%), even though we observed sources with an arcsecond core flux density as low as 5 mJy at 5 GHz. This result
confirms the presence of compact nuclei at the center of radio galaxies with very low power radio core.\\
b) The one-sided jet morphology is the predominant structure on the parsec-scale (in $\sim$ 80\% of sources). 
This is in agreement with a random orientation of radio sources and a high jet velocity ($\beta\sim$ 0.9). No intrinsic difference in jet velocity and morphology has been found between high and
low power radio galaxies.\\
c) With very few exceptions, the parsec and kiloparsec-scale radio structures are aligned confirming that the large bends present in some BL Lacs are likely amplified by the small jet
orientation angle with respect to the line-of-sight. In sources with aligned pc and kpc scale structure,
the main jet is always on the same side with respect to the nuclear emission.\\
d) We find two sources with a Z-shaped structure on the pc-scale (4c26.42, Liuzzo et al. 2009a, and 3C310, Liuzzo et al. 2009b) suggesting the presence of low velocity jets in these objects.\\
e) In $\sim$ 40\% of the sources, there is evidence of nuclear variability and/or of a significant
sub-kpc-scale structure which will be better investigated with the EVLA at high frequency or with
the E-MERLIN array. \\
f) $\sim$3\% sources present very high core dominance values ($>$10) which suggest the presence of recurrent or re-starting activity. \\
g) In 20/94 objects (mainly among these 17 recently observed very faint BCS sources), there is evidence of low activity state, but not always completely quiescent. These sources show in fact low core
dominance values ($<$0.25) but the presence of pc-scale core and in some cases even the jet.

\section{Future work}
The analysis of these new VLBI data and the statistical study of the BCS are still in progress. In particular, we have to deeply investigate our preliminary results to properly address the points discussed in Sect. 2. Our final aim is to discuss more in general the BCS nuclear multiband properties in comparison with those of low-z BL Lac objects (Giovannini et al. 2014, Liuzzo et al. 2013) for a better understanding of unified models of radio loud AGN.\\
{\bf Acknowledgments:} We sincerely thank the organizers for this very interesting meeting. 
\\\\
{\bf References.}\\
Giovannini G. et al. 2001: ApJ 552, 508          \\  
Giovannini G. et al. 2005: ApJ 618, 635  	\\
Giovannini G. et al. 2014: 2014cosp...40E.998G  \\
Kharb P. et al. 2010:  ApJ 710, 764K\\
Liuzzo E. et al. 2009a: A\&A 501, 933\\
Liuzzo E. et al. 2009b: A\&A 505, 509\\
Liuzzo et al. 2013:  A\&A 560, 23L\\
Taylor, G.B. et al. 1994: ApJSS 95, 345.

\begin{figure}
\centering
\includegraphics[width=16cm, angle=0]{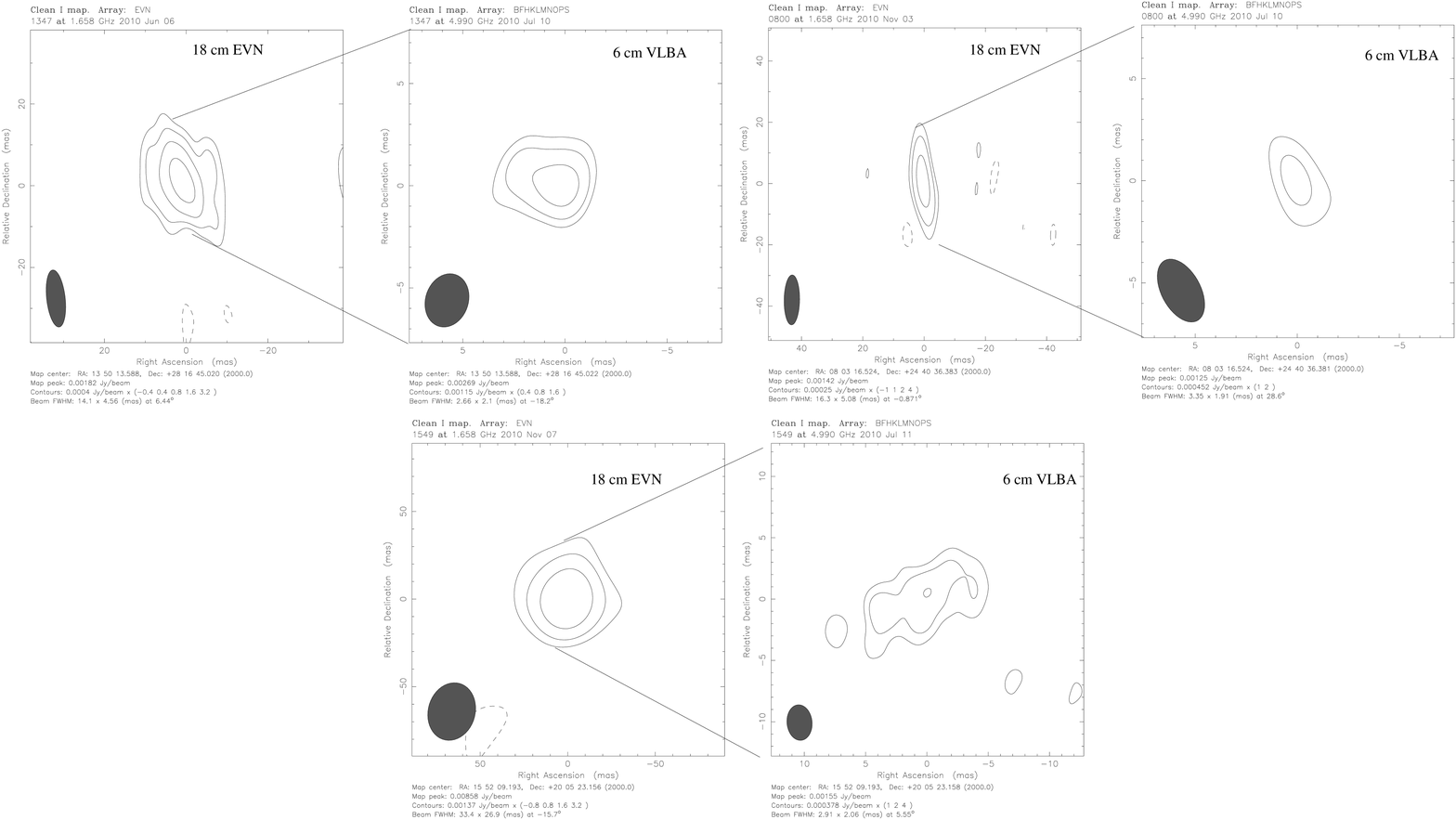}
\caption{\footnotesize {18 cm EVN and 6 cm VLBA images of 3 of the faintest BCS sources recently observed.}}
\label{fig_images}
\end{figure}
\end{document}